\documentclass[12pt]{iopart}

\usepackage{graphicx,bbm,amssymb,amsopn}

\newcommand{\un}[1]{{\underline{#1}}}
\newcommand{\bra}[1]{{\langle #1 \vert}}

\newcommand{\ket}[1]{{\vert #1 \rangle}}
\newcommand{\xket}[1]{{\vert #1 \rangle}}

\newcommand{\xbraket}[2]{\langle #1 \vert #2 \rangle}

\newcommand{\ave}[1]{{\langle #1\rangle}}
\newcommand{\rs}{ {\rm s} }
\newcommand{\rb}{ {\rm b} }

\newcommand{\ii}{ {\rm i} }
\newcommand{\dd}{ {\rm d} }

\newcommand{\CC}{\mathbb{C}}
\newcommand{\cc}{ {\hat c} }
\newcommand{\bb}{ {\hat b} }
\newcommand{\aaa}{ {\hat a} }
\newcommand{\y}{{\rm y}}
\newcommand{\x}{{\rm x}}
\newcommand{\z}{{\rm z}}
\newcommand{\LL}{{\hat {\cal L}}}
\newcommand{\DD}{{\hat {\cal D}}}
\newcommand{\NN}{{\hat {\cal N}}}
\newcommand{\PP}{{\hat {\cal P}}}
\newcommand{\mm}[1]{{\mathbf{#1}}}

\def\L{{\rm L}}
\def\R{{\rm R}}
\def\tr{{\,{\rm tr}\,}}
\def\ad{{\,{\rm ad}\,}}
\def\one{\mathbbm{1}}
\def\re{{\,{\rm Re}\,}}

\def\ness{\xket{{\rm NESS}}}

\begin{document}

\title[Exact solution of Markovian master equations for quadratic fermi systems]{Exact solution of Markovian master equations for quadratic fermi systems:  thermal baths, open XY spin chains, and non-equilibrium phase transition}

\author{Toma\v{z} Prosen and Bojan \v Zunkovi\v c}

\address{Department of physics, FMF, University of Ljubljana,
Jadranska 19, SI-1000 Ljubljana, Slovenia}

\date{\today}

\begin{abstract}
We generalize the method of {\em third quantization} to a unified exact treatment of Redfield and Lindblad 
master equations for open quadratic systems of $n$ fermions in terms of diagonalization of $4n\times 4n$ matrix.
Non-equilibrium thermal driving in terms of the Redfield equation is analyzed in detail.  We explain how to compute all physically relevant quantities, such as non-equilibrium expectation values of local observables, various entropies or information measures, or time evolution and properties of relaxation. 
We also discuss how to exactly treat explicitly time dependent problems.
The general formalism is then applied to study a thermally driven open XY spin 1/2 chain.
We find that recently proposed non-equilibrium quantum phase transition in the open XY chain survives the thermal driving within the Redfield model. In particular, the
phase of long-range magnetic correlations can be characterized by hypersensitivity of the non-equilibrium-steady state to external (bath or bulk) parameters.
Studying the heat transport we find negative thermal conductance for sufficiently strong thermal driving, as well as non-monotonic dependence of the heat current on the strength 
of the bath coupling.
\end{abstract}


\maketitle

\section{Introduction}

One of the main challenges of the many-body theory and non-equilibrium statistical mechanics is to understand the properties of relaxation of large interacting quantum systems. There are two common approaches to this type of problems. One important direction is to try to define dynamics in the thermodynamic limit and to investigate its properties with rigorous mathematical methods of operator algebras \cite{araki,ruelle,jaksic}. However, in this context explicit results which go beyond existence proofs are quite limited. A second approach is to split a large system into a tensor product of a smaller system of interest, and the rest (environment), and trying to eliminate all the degrees of freedom of the large, macroscopic environment (see e.g. \cite{petruccione,alicki}). This approach, although involving a series of approximations, is usually more fruitful for explicit calculations and quantitative analyses.  We may be interested either in relaxation to equilibrium or non-equilibrium steady states, depending on the equal or non-equal 
values of thermodynamic potentials assigned to possibly several pieces of environment - which we shall call {\em the baths}.
Such calculations of the quantitative properties of steady states may be very useful, for example in the realm of transport theory \cite{wich} as may complement the linear response calculations and suggest non-linear response or far-from-equilibrium effects.

However, to date we have had a very few explicit calculations of non-equilibrium properties of open many body quantum systems, and mainly they had to focus on small systems with a 
single or a pair of degrees of freedoms (such as spins, or bosons), see for example \cite{burgarth,ban}.
The reason is that there has been no theoretical techniques to deal with open many-body problems except for the Keldysh formalism of non-equilibrium Green's functions, which however can easily get too involved for explicit calculations. Recently, two new directions have been proposed, both in the direction of numerical simulation and theoretical analysis. Namely, in the context of numerical simulations of open many-body systems, time-dependent density matrix renormalization group techniques \cite{dmrg} have been demonstrated to efficiently simulate relaxation to steady states with the Lindblad master equation \cite{pz09}. On the other hand, it has been shown \cite{njp} that the Lindblad equation for general quadratic fermionic systems, for example for XY-like quantum spin chains which are mappable to quadratic fermionic systems, can be solved explicitly with the technique of canonical quantization in the Fock space of operators -  {\em third quantization} for short.

In this paper we shall show how the third quantization can be generalized to treat quadratic systems with arbitrary Markovian master equations , which is not necessarily of the Lindblad form. In particular, we shall focus on the Redfield dissipator in terms of which we can simulate simple thermal reservoirs, and thermal driving of the system under non-equilibrium conditions. After giving a short account on mathematical formulation of
Markovian master equations and the basic physical assumptions and approximations involved in the derivation - in section \ref{sect:markovian}, we shall in section \ref{sect:method} present a short but self-contained generalization of the theory \cite{njp}. In addition, we shall outline the calculation of
dynamical correlation functions in Liouvillean dynamics, and formulate an exact treatment of explicitly time-dependent quantum Liouville problems.
In section \ref{sect:xy} we shall apply our technique to treat an open XY spin chain in the non-equilibrium Redfield model. We shall outline several intriguing exact numerical results on
large spin chains. In particular, we show that recently announced quantum phase transition in the open XY chain in the local Lindblad bath model, generalizes also to non-equilibrium thermal Redfield model with qualitatively identical characteristics. The transition is characterized by spontaneous emergence of long range magnetic correlations, and hypersensitivity of the steady state to external system's parameters, when the transverse magnetic field drops bellow the critical value $|h| < h_{\rm c} = |1-\gamma^2|$ where $\gamma$ is the anisotropy parameter. Furthermore, we analyze in 
some detail the heat transport in XY chain, and find regions of {\em negative differential heat conductance} for strong thermal driving, namely non-monotonic dependence of the heat current on the
temperature difference between the baths.

\section{Markovian master equations in non-equilibrium quantum physics}

\label{sect:markovian}

Decomposing the Hilbert space of the universe into a tensor product ${\cal H} = {\cal H}_\rs \otimes {\cal H}_\rb$ of the {\em central system} ${\cal H}_\rs$ and the {\em bath} (or a set of baths) ${\cal H}_\rb$ (environment), one writes the total Hamiltonian as
\begin{equation}
H = H_\rs \otimes \one_\rb + \one_\rs \otimes H_\rb + \lambda \sum_\mu X_\mu\otimes Y_\mu,
\end{equation}
where $X_\mu$, are linear operators over ${\cal H}_\rs$, and $Y_\mu$ linear operators over ${\cal H}_\rb$.
Note that $X_\mu,Y_\mu$ can always be chosen to be Hermitian, so this shall be assumed throughout this paper. The Markovian quantum master equation for the time evolution of the central systems's density matrix $\rho(t)$ is derived \cite{petruccione} using three main assumptions: (i) weak coupling (assuming $\lambda$ to be small), (ii) factorizability of the initial density matrix $\rho_\rs(0)\otimes\rho_\rb(0)$, and (iii) Born-Markov approximation which rests upon the assumption that the bath-correlation functions
\begin{equation}
\Gamma^\beta_{\mu,\nu}(t) := \lambda^2 \tr(\tilde{Y}_\mu(t) Y_\nu e^{-\beta H_\rb})/\tr e^{-\beta H_\rb}, \quad \tilde{Y}_\mu(t):=e^{\ii t H_\rb} Y_\mu e^{-\ii t H_\rb}
\label{eq:spectf}
\end{equation}
decay on much shorter time scale than the central systems dynamics $\tilde{X}_\mu(t):=e^{\ii t H_\rs} X_\mu e^{-\ii t H_\rs}$. We use units in which Planck's constant $\hbar=1$, and may use different inverse temperatures $\beta$ for different pieces of the environment (for different baths).
The resulting master equation is referred to as the Redfield equation
\begin{equation}
\frac{\dd}{\dd t}\rho(t) = -\ii [H_\rs,\rho(t)] + \hat{\cal D}\rho(t),
\label{eq:master}
\end{equation}
where the dissipator-map has a memoryless kernel with the following general form
\begin{equation}
\hat{\cal D}\rho= \sum_{\mu,\nu}\int_0^\infty\dd\tau \Gamma_{\nu,\mu}^\beta(\tau)[\tilde{X}_\mu(-\tau)\rho,X_\nu] + h.c.
\label{eq:redfD}
\end{equation}
If one additionally assumes the so-called rotating wave-approximation, one arrives at the dynamical semi-group which manifestly preserves the positivity of density matrix at all times\footnote{This is not the case for equation (\ref{eq:master},\ref{eq:redfD}) which allows for possible breaking of positivity at initial short time interval, the so called {\em sleapage time}.} and can be generally described by the dissipator in the Lindblad form
\begin{equation}
\hat{\cal D}'\rho = \sum_{\mu,\nu} \gamma_{\nu,\mu}[X_\mu \rho,X_\nu] + h.c. ,
\label{eq:lindD}
\end{equation}
where the only condition is that $\mm{\gamma}$ is a Hermitian  $\gamma_{\mu,\nu}=\gamma^*_{\nu,\mu}$ and positive definite matrix.
The standard Lindblad form is obtained by diagonalizing the matrix $\mm{\gamma}$ whose eigenvectors yield the usual Lindblad operators.
The important property of the bath-correlation functions (\ref{eq:spectf}) (which constitute all that we need to know about the baths) is the 
Kubo-Martin-Schwinger({\em KMS}) condition
\begin{equation}
\Gamma^\beta_{\mu,\nu}(-t-\ii\beta) = \Gamma^\beta_{\nu,\mu}(t),
\label{eq:KMS}
\end{equation}
which is needed to prove that the thermal state $\rho_{\rm gibbs} = e^{-\beta H_\rs}/\tr e^{-\beta H_\rs}$ is a steady state of  the master equation (\ref{eq:master}),
provided that all baths are thermalized to the {\em same} inverse temperature\footnote{With an additional technical condition of neglecting the Cauchy principal value 
contribution to the time integral \ref{eq:redfD}, see the discussion at the end of subsection \ref{bilinear}}.
However, in case of several thermal baths with possibly different temperatures we may expect that $\rho(t)$ relaxes to a
physically very interesting {\em non-equilibrium-steady-state} (NESS).

\section{Diagonalization of quantum Liouvilleans for quadratic fermi systems}

\label{sect:method}

In this section we give a short account on the general technique of canonical quantization in the Liouvile space (`third quantization') and complete diagonalization
of Markovian master equations (\ref{eq:master}), with (\ref{eq:redfD}) or (\ref{eq:lindD}), for quadratic fermionic problems.
We treat a {\em finite} problem with $n$ fermionic degrees of freedom, described by $2n$ anti-comuting Hermitian operators
$w_j$, $j=1,2,\ldots,2n$,  $\{w_j,w_k\} = 2\delta_{j,k} $, in which the Hamiltonian $H$ may take a general quadratic form and the coupling operators may be general linear forms:
\begin{eqnarray}
H_{\rm s} &=& \sum_{j,k=1}^{2n} w_j H_{j,k} w_k = \un{w} \cdot \mm{H}\, \un{w}, \label{eq:hamil}  \\
X_\mu &=& \sum_{j=1}^{2n} x_{\mu,j} w_j = \un{x}_\mu \cdot \un{w}\ .  \label{eq:lindb}
\end{eqnarray}
Thus, $2n \times 2n$ matrix $\mm{H}$ can be chosen to be antisymmetric $\mm{H}^T = -\mm{H}$. 
Throughout this paper $\un{x}=(x_1,x_2,\ldots)^T$ will designate a vector (column) of appropriate scalar valued or operator valued symbols $x_k$.
This formalism is immedately applicable either for describing, (i) {\em physical fermions}  $c_m$, $m=1,2,\ldots, n$,
where $w_{2m-1}= c_m + c_m^\dagger$, $w_{2m} = \ii(c_m- c^\dagger_m)$, or (ii) XY-like systems of spins $1/2$ with canonical Pauli operators $\vec{\sigma}_m$, $m=1,2,\ldots, n$,
where the fermionic operators are represented by the famous Jordan-Wigner transformation
\begin{equation}
w_{2m-1} = \sigma^\x_m \prod_{m'<m} \sigma^\z_{m'}\, , \qquad
w_{2m} = \sigma^\y_m \prod_{m'<m} \sigma^\z_{m'} \,.
\label{eq:jordan}
\end{equation}

\subsection{Fock space of operators}

The fundamental concept for our analysis is a Fock space structure over the $4^n$ dimensional Liouville space of operators ${\cal K}$, which density matrix $\rho(t)$ is also a member of. From here on, we shall adopt Dirac bra-ket notation for the operator space ${\cal K}$ which is fixed by the following definition of the inner product
\begin{equation}
\xbraket{x}{y} = \tr x^\dagger y, \qquad x,y \in {\cal K}.
\end{equation}
We note that $2^{2n}$ operator-products $\xket{P_{\un{\alpha}}}$, labelled with a binary multi-index $\un{\alpha}$
\begin{equation}
P_{\alpha_1,\alpha_2,\ldots,\alpha_{2n}} := 2^{-n/2} w_1^{\alpha_1}w_2^{\alpha_2}\cdots w_{2n}^{\alpha_{2n}}, \qquad
\alpha_j\in\{ 0,1\}
\label{eq:defP}
\end{equation}
constitute a complete {\em orthonormal} basis of ${\cal K}$ with respect to an inner product.

In fact it is easy to show that $\xket{P_{\un{\alpha}}}$ is a fermionic Fock basis, and powers $1$ in the product (\ref{eq:defP}) can be considered like
a sort of Fermionic excitations, if we define the following set of linear {\em annihilation maps} $\hat{c}_j$ over\footnote{We shall use notation where linear maps over the operator
space (in physics literature sometimes referred to as ``super-operators") are designated by $\hat{}$.} ${\cal K}$
\begin{equation}
\cc_j \xket{P_{\un{\alpha}}} = \alpha_j \xket{w_j P_{\un{\alpha}}},
\label{eq:defanih}
\end{equation}
and derive the actions of their Hermitian adjoints - the {\em creation linear maps} $\hat{c}^\dagger$,
\begin{equation}
\cc^\dagger_j \xket{P_{\un{\alpha}}} = (1-\alpha_j) \xket{w_j P_{\un{\alpha}}},
\end{equation}
which satisfy canonical anticommutation relations
\begin{equation}
\{\cc_j,\cc_k\} = 0, \qquad \{\cc_j,\cc_k^\dagger\} = \delta_{j,k}, \qquad j,k=1,2,\ldots, 2n.
\end{equation}

\subsection{Bilinear form of the Liouvillean}

\label{bilinear}

The aim is now to show that the generator of the master equation (\ref{eq:master}) 
\begin{equation}
\LL := -\ii\ad H + \DD
\label{eq:Liouvillean}
\end{equation} is in general a {\em quadratic form} in these {\em adjoint fermionic
maps} $\hat{c}_j,\hat{c}^\dagger_j$. In order to see that clearly, let us define the {\em left} and {\em right} multiplication maps over ${\cal K}$
\begin{equation}
\hat{w}^\L_j \ket{x} := \ket{w_j x},\qquad \hat{w}^\R_j\ket{x} := \ket{x w_j}.
\end{equation}
Inspecting the actions of $\hat{w}^\L_j,\hat{w}^\R_j$ on the Fock basis $\xket{P_{\un{\alpha}}}$ one arrives at the following useful identities
\begin{eqnarray}
\hat{w}^\L_j &=& \hat{c}_j + \hat{c}_j^\dagger, 
\label{eq:defWL}\\
\hat{w}^\R_j &=& \PP (\hat{c}_j - \hat{c}_j^\dagger) =  -(\hat{c}_j - \hat{c}_j^\dagger)\PP,
\label{eq:defWR}
\end{eqnarray}
where 
\begin{equation}
\PP := \exp(\ii \pi \NN), {\quad \rm and\quad} \NN := \sum_{j=1}^{2n} \hat{c}^\dagger_j \hat{c}_j
\end{equation} 
are a {\em parity map}, and a {\em number map}, respectively, which count the parity and number of the adjoint fermionic excitations (number of factors in (\ref{eq:defP})).
Note that $\PP$, anticommutes with all $\hat{c}_j,\hat{c}^\dagger_j$, hence the second equality of (\ref{eq:defWR}), and $\PP^2 = \hat{\one}$.

The unitary part of the Liouvillean (\ref{eq:Liouvillean}) now trivially reads
\begin{equation}
-\ii \ad H_{\rm s} = -\ii\un{\hat{w}}^\L \cdot \mm{H} \un{\hat{w}}^\L + \ii \mm{H}\un{\hat{w}}^\R \cdot \un{\hat{w}}^\R = -4\ii \un{\hat c}^\dagger\cdot\mm{H}\un{\hat c} \,. 
\label{eq:unit}
\end{equation}

The dissipator (\ref{eq:redfD}) can be represented as a map over ${\cal K}$ as
\begin{equation}
\!\!\!\!\!\!\!\!\!\!\!\DD =
\sum_{\mu,\nu}\sum_{j,k=1}^{2n} x_{\nu,k}
\int_0^\infty\!\!\!\dd\tau f_{\mu,j}(-\tau) \left(
\Gamma_{\nu,\mu}^\beta(\tau)\LL'_{j,k} +
\Gamma_{\nu,\mu}^{\beta *}(\tau) \LL''_{j,k}\right),
\label{eq:DD}
\end{equation}
where $\un{f}_\mu(t)$ is a (real-valued) propagator of Heisenberg dynamics in the closed system
\begin{equation}
\tilde{X}_\mu(t) = \un{x}_\mu\cdot \exp(-\ii \ad H_{\rm s} t) \un{w} =: \un{f}_\mu(t)\cdot \un{w},
\end{equation}
which - due to (\ref{eq:unit}) - can be explicitly solved for a quadratic Hamiltonian (\ref{eq:hamil}), giving
\begin{equation}
\un{f}_\mu(t) = \exp(4\ii\mm{H} t) \un{x}_\mu,
\label{eq:heis}
\end{equation}
and
\begin{equation}
\LL'_{j,k}\ket{x} := \ket{[w_j x,w_k]},\quad
\LL''_{j,k}\ket{x} := \ket{[w_k,x w_j]}
\end{equation}
are fundamental basis dissipators which using (\ref{eq:defWL},\ref{eq:defWR}) evaluate to
\begin{eqnarray}
\!\!\!\!\!\!\!\!\!\!\LL'_{j,k} &=& \hat{w}^\L_j \hat{w}^\R_k - \hat{w}^\L_k \hat{w}^\L_j 
= (\hat{\one}+\PP)(\hat{c}^\dagger_j \hat{c}^\dagger_k - \hat{c}^\dagger_k \hat{c}_j) 
+ (\hat{\one}-\PP)(\hat{c}_j \hat{c}_k - \hat{c}_k \hat{c}^\dagger_j), \label{eq:defL1} \\
\!\!\!\!\!\!\!\!\!\!\LL''_{j,k} &=& \hat{w}^\L_k \hat{w}^\R_j - \hat{w}^\R_k\hat{w}^\R_j 
= (\hat{\one}+\PP)(\hat{c}^\dagger_k \hat{c}^\dagger_j - \hat{c}^\dagger_k \hat{c}_j) 
+ (\hat{\one}-\PP)(\hat{c}_k \hat{c}_j - \hat{c}_k \hat{c}^\dagger_j). \label{eq:defL2}
\end{eqnarray}
It will prove useful if we express the internal dynamics (\ref{eq:heis}) explicitly in terms of eigenvalues and eigenvectors of the Hamiltonian matrix $\mm{H}$.
Since $2n\times 2n$ matrix is anti-symmetric and Hermitian, its real eigenvalues come in pairs $\epsilon_m,-\epsilon_m,j=1,\ldots,n$,
with the corresponding eigenvectors $\un{u}_m,\un{u}_m^*$, namely $\mm{H}\un{u}_m = \epsilon_m \un{u}_m$ and $\mm{H}\un{u}^*_m = -\epsilon_m \un{u}^*_m$ since 
$\mm{H}^*=-\mm{H}$.
The eigenvectors may and should always be chosen orthonormal (even in the case of degeneracies), meaning 
\begin{equation}
\un{u}_l\cdot\un{u}_m = 0, \qquad \un{u}_l\cdot\un{u}^*_m =\delta_{l,m}.
\end{equation}
Then the spectral decomposition of the Heisenberg dynamics reads
\begin{equation}
\un{f}_\mu(t) = \sum_{m=1}^n \left( e^{-4\ii \epsilon_m t} (\un{x}_\mu\cdot \un{u}_m) \un{u}^*_m + e^{4\ii \epsilon_m t} (\un{x}_\mu\cdot \un{u}^*_m) \un{u}_m\right).
\end{equation}

Note that $\PP_\pm=(\hat{\one} \pm \PP)/2$ are {\em orthogonal projectors} which commute with {\em all} the terms (\ref{eq:unit},\ref{eq:defL1},\ref{eq:defL2}) that constitute the Liouvillean (\ref{eq:Liouvillean}), $[\PP_\pm,\LL]=0$, and hence the dynamics (\ref{eq:master}) does not mix the operator subspaces ${\cal K}^\pm = \PP_\pm{\cal K}$
composed of even/odd number of fermionic operators. 
Since we are mainly interested in expectation values of even observables, such as currents and densities, we shall in the present paper focus on the dynamics in
the subspace ${\cal K}^+$ only, and consider the corresponding Lioivillean $\LL|_{{\cal K}^+}$
\begin{equation}
\LL_+ = \PP_+ \LL \PP_+.
\end{equation}
The extension to the odd parity subspace is straightforward.
Collecting the results (\ref{eq:unit},\ref{eq:DD},\ref{eq:defL1},\ref{eq:defL2}) it is now obvious that $\LL_+$ is a bilinear form in $\hat{c}^\dagger_j$ and $\hat{c}_j$.
For convenience, we define $4n$ Hermitian Majorana maps $\hat{a}_r, r=1,\ldots 4n$
\begin{equation}
\hat{a}_{2j-1} = \frac{1}{\sqrt{2}}(\hat{c}_j + \hat{c}^\dagger_j),\quad 
\hat{a}_{2j} = \frac{\ii}{\sqrt{2}}(\hat{c}_j - \hat{c}^\dagger_j),
\end{equation}
and express the Liouvillean as
\begin{equation}
\LL_+ = \un{\hat a}\cdot \mm{A}\un{\hat a} - A_0 \hat{\one},
\label{eq:LiouvA}
\end{equation}
where the $4n\times 4n$ complex antisymmetrix matrix $\mm{A}$, later referred to as a {\em structure matrix}, and a scalar $A_0$, can be expressed as  
\begin{eqnarray}
A_{2j-1,2k-1} &=&-2\ii H_{j,k}-M_{j,k}+M_{k,j} \small,\nonumber \\
A_{2j-1,2k}   &=& \;\;\;\; \ii M_{k,j} + \ii M_{j,k}^* \small,\nonumber \\
A_{2j,2k-1}   &=& -\ii M_{j,k} - \ii M_{k,j}^* \small,\nonumber \\
A_{2j,2k}     &=&-2\ii H_{j,k}-M^*_{j,k}+M^*_{k,j}\small, \label{eq:explA}\\
A_0 &=& \tr\mm{M} + \tr\mm{M}^*\small, \nonumber
\end{eqnarray}
where $\mm{M}$ is a $2n \times 2n$ bath-matrix which can be compactly written as
\begin{eqnarray}
\mm{M} &:=& \sum_{\nu} \un{x}_\nu \otimes \un{z}_{\nu}, \\
\un{z}_\nu &:=& \sum_\mu\int_0^\infty \dd\tau \Gamma^\beta_{\nu,\mu}(\tau)\un{f}_\mu(-\tau). \label{eq:zdef}
\end{eqnarray}
Defining the bath-spectral functions
$\tilde{\Gamma}^\beta_{\mu,\nu}(\omega) := \frac{1}{2\pi}\int_{-\infty}^\infty \!\!\dd t\, \Gamma^\beta_{\mu,\nu}(t) e^{-\ii \omega t}
$ for which the KMS condition reads
\begin{equation}
\tilde{\Gamma}_{\mu,\nu}^\beta(-\omega) = e^{\beta\omega} \tilde{\Gamma}^\beta_{\nu,\mu}(\omega),
\end{equation}
and extending the range of integration in (\ref{eq:zdef}) to $[-\infty,\infty]$, or better to say, neglecting the Cauchy principal value parts in the integrals - which exactly amounts to neglecting the {\em Lamb-shift Hamiltonian term} \cite{petruccione} in the master equation  - we obtain a very simple expression (involving only finite sums) for the bath-vectors
\begin{equation}
\un{z}_\nu = \pi \sum_\mu\sum_{m=1}^n  \tilde{\Gamma}^\beta_{\nu,\mu}(4\epsilon_m)\left((\un{x}_\mu\cdot \un{u}^*_m)\un{u}_m
+  e^{4\epsilon_m \beta}(\un{x}_\mu\cdot \un{u}_m)\un{u}^*_m\right).
\label{eq:znu}
\end{equation}

At this point a remark on neglecting the Lamb-Shift term is in order. As the Redfield model already involves a series of physical assumptions and approximations it is somewhat difficult to argue under what conditions these terms can be dropped on the same level of approximations. However, one can straightforwardly show using the KMS condition (\ref{eq:KMS}) and Hermiticity
$(\Gamma_{\mu,\nu}^\beta(\tau))^* = \Gamma^\beta_{\nu,\mu}(\tau)$ that {\em only if} the Cauchy principal value terms are dropped (i.e. if the range of integration in (\ref{eq:redfD}) is extended
to $[-\infty,\infty]$) the Redfield dissipator annihilates the Gibbs state $\hat{\cal D}\ket{ e^{-\beta H_{\rm s}}} = 0$, and hence Gibbs state is the steady state of equilibrium thermal Redfield model.

Note again that the inverse temperature in (\ref{eq:znu}) could in principle be a function of the bath-index $\beta=\beta_\nu$ in case one would be interested in non-equilibrium situation with couplings to several different temperatures. But we should stress that {\em different} temperatures only make sense among {\em uncorrelated baths} for which 
$\Gamma^\beta_{\mu,\nu} \equiv 0$ for any $\beta$.

We note also that the present formalism uniformly covers both the Redfield and the Lindblad master equations, as the Lindblad dissipator (\ref{eq:lindD}) is obtained from
(\ref{eq:redfD}) by simply taking the limit $\Gamma^\beta_{\mu,\nu}(t) = \gamma_{\mu,\nu} \delta(t+0)$, and then the bath-matrix reduces to a Hermitian form
$\mm{M} = \sum_{\nu,\mu} \gamma_{\nu,\mu} \un{x}_\nu \otimes \un{x}_\mu = \mm{M}^\dagger$ which is equivalent to the one used in \cite{njp}.

\subsection{Static Liouvillean: normal modes, non-equilibrium steady state and decay spectrum}

\label{subsect:static}

Having the compact form of the Liouvillean  (\ref{eq:LiouvA})  -- and assuming for the time being that the structure matrix $\mm{A}$ is {\em static} i.e. there is {\em no explicit time dependence} in   the matrix $\mm{H}$ or coupling vectors $\un{x}_\mu$ -- we follow Ref.\cite{njp} and explicitly construct its normal form, the NESS which is exactly the right-vacuum state of 
(\ref{eq:LiouvA}) $\LL_+ \ness= 0$, the spectral gap, and the full spectrum of Liouvillean decay modes, all in terms of spectral decomposition of $4n\times 4n$ matrix
$\mm{A}$. We state the main results here in a compact form.

Assuming the structure matrix is {\em diagonalizable}, its eigenvalues  can be paired as $\beta_j,-\beta_j,j=1,\ldots,2n$, assuming $\re\beta_j \ge 0$, 
and its eigenvectors $\un{v}_{2j-1}$ (corresponding to $\beta_j$), and $\un{v}_{2j}$ (corresponding to $-\beta_j$) can always be normalized -- irrespective of possible degeneracies of
among $\beta_j$, which shall be called {\em rapidities} -- such that 
\begin{equation}
\mm{V}\mm{V}^T = \mm{J},\quad \mm{J} := \sigma^{\x}\otimes \one_{2n} =
\pmatrix{0 & 1 & 0 & 0 & \cdots \cr
                1 & 0 & 0 & 0 & \cdots \cr
                0 & 0 & 0 & 1 & \cdots \cr
                0 & 0 & 1 & 0 & \cdots \cr
\vdots &\vdots & \vdots & \vdots& \ddots \cr},
\label{eq:norm}
\end{equation}
where $\mm{V}$ is $4n\times 4n$ matrix whose $r$th row is given by $\un{v}_r$,
$V_{r,s} := v_{r,s}$.
Thus the structure matrix allows the following decomposition
\begin{equation}
\mm{A} = \mm{V}^T {\rm diag}\{\beta_1,-\beta_1,\ldots,\beta_{2n},-\beta_{2n}\}\mm{J}\mm{V},
\end{equation}
which after plugging into the Liouvillean (\ref{eq:LiouvA}) immediately brings it to a {\em normal form}
\begin{equation}
\LL_+ = -2\sum_{j=1}^{2n} \beta_j \bb'_j \bb_j\small,
\end{equation}
where 
\begin{equation}
\hat{b}_j := \un{v}_{2j-1}\cdot\un{\hat a}\small,\quad  \hat{b}'_j := \un{v}_{2j}\cdot\un{\hat a}\small,
\end{equation}
are the {\em normal-master-mode} (NMM) maps, satisfying almost canonical anti-commutation relations
\begin{equation}
\{\bb_j,\bb_k\} = 0\small, \qquad \{\bb_j,\bb'_k\} = \delta_{j,k}\small, \qquad \{\bb'_j,\bb'_k\}=0 \small.
\label{acar}
\end{equation}
The map $\bb_j$ could be interpreted as an annihilation map and $\bb'_j$ as a
creation map of $j$th NMM, but we should note that $\bb'_j$ is in general {\em not} the
Hermitian adjoint of $\bb_j$. The right-vacuum is now essentially defined by $\bb_j\ness = 0$, whereas the left-vacuum is trivial $\bra{1}\LL_+ = 0$ and satisfies
$\bra{1}\bb'_j = 0$.

Assuming that the whole rapidity spectrum is strictly away from the real line $\re\beta_j > 0$, we state the following exact results:
\begin{enumerate}
\item $\ness$ is unique.
\item Almost any initial density matrix relaxes to NESS with an exponential rate $\Delta =  2\min \re\beta_j$ (the spectral gap of the Liouvillean).
The complete spectrum of $4^n$ eigenvalues of $\LL_+$ is obtained by all possible binary linear combinations
$\lambda_{\un{\nu}} = -2\un{\nu}\cdot\un{\beta}$, $\nu_j\in\{0,1\}$.
\item The expectation value of {\em any quadratic observable} $w_j w_k$  in a (unique) NESS can be explicitly 
computed as
\begin{eqnarray}
\ave{w_j w_k}_{\rm NESS} &:=& \tr w_j w_k \rho_{\rm NESS} = 2 \bra{1}\hat{a}_{2j-1} \hat{a}_{2k-1}\ness 
\label{eq:line1} \\
&=&  2\sum_{m=1}^{2n} 
v_{2m,2j-1} v_{2m-1,2k-1}  \label{eq:wwv} \\
&=&  -\frac{1}{\pi}\int_{-\infty}^\infty\! {\rm d}\omega\, G_{2j-1,2k-1}(\omega), \label{eq:wwg}
\end{eqnarray}
where
\begin{equation}
\mm{G}(\omega) := (\mm{A} - \ii \omega \one)^{-1}
\label{eq:Green}
\end{equation}
is a matrix of the non-equilibrium Green's function. The first equality is proven in \cite{njp} \footnote{Small simplification has been made with respect to the statement of Theorem 3 of Ref.\cite{njp} which has been pointed out by I. Pi\v zorn \cite{iztok}.} whereas the last equality requires a simple contour integration on the spectral decomposition
of the resolvent (\ref{eq:Green}).
\item The Wick theorem may be used for calculation of expectation values of arbitrary higher order (even!) observables by sums of all possible pairwise contractions of
the form (\ref{eq:line1}).
\end{enumerate}
Note that as soon as some of the rapidities condense to the imaginary axis, or vanish, NESS typically becomes non-unique (see Ref. \cite{pp09} for a detailed discussion
of Liouvillean degeneracies).

\subsection{Static Liouvillean: time-dependent correlation functions}

The complete Liouvillean propagator can be written explicitly as
\begin{equation}
\!\!\!\!\!\!\!\!\!\!\!\!\!\!\!\!\!\!\!\!\!\!\!\exp(t \LL_+) = \sum_{\un{\nu} \in \{0,1\}^{2n}} \exp(-2t \un{\nu}\cdot\un{\beta}) (\bb'_{1})^{\nu_1}\cdots (\bb'_{2n})^{\nu_{2n}} \ness\bra{1}(\bb_{2n})^{\nu_{2n}}\cdots(\bb_1)^{\nu_1}.
\label{eq:prop}
\end{equation}
It may be of some physical interest to evaluate dynamical response after perturbing the NESS by multiplying it with some local observable.
In order to avoid discussion of negative parity dynamics $\LL_-$ we take a pair of simplest even-order, quadratic observables, and define the corresponding non-equilibrum time-dependent correlation function - or non-equlibrium response function - as
\begin{eqnarray}
C_{(j,k),(l,m)}(t) &:=&\ave{w_j(t)w_k(t) w_l(0)w_m(0)}_{\rm NESS} = \nonumber \\
&=& 4 \bra{1}\aaa_{2j-1}\aaa_{2k-1} \exp(t\LL_+)\aaa_{2l-1}\aaa_{2m-1}\ness.
\end{eqnarray}
Expressing the multiplication maps $\aaa_{2j-1} = \sum_{r=1}^{2n}(V_{2r,2j-1} \bb_r + V_{2r-1,2j-1} \bb'_r)$ and 
plugging in the propagator (\ref{eq:prop}), while noting that only the terms with 0 or 2 Liouvillean excitations contribute, we obtain a simple expression
\begin{eqnarray}
C_{(j,k),(l,m)}(t) &=& 4\left(\sum_{r=1}^{2n} v_{2r,2j-1}v_{2r-1,2k-1}\right)\left( \sum_{r'=1}^{2n} v_{2r',2l-1}v_{2r'-1,2m-1}\right) \nonumber \\
&+&4\!\!\sum_{1\le r < r'\le 2n}\!\!\!
e^{-2t(\beta_{r}+\beta_{r'})}
\left(v_{2r',2j-1}v_{2r,2k-1} \!-\! v_{2r,2j-1}v_{2r',2k-1}\right) \nonumber\\
&&\qquad\quad\times\left(v_{2r'-1,2l-1}v_{2r-1,2m-1} \!-\! v_{2r-1,2l-1}v_{2r'-1,2m-1}\right).
\end{eqnarray}

\subsection{Time-dependent Liouvilleans}

In this subsecton we indicate how to efficiently treat explicitly time-dependent master equations, written in third quantized form as
\begin{equation}
\frac{\dd}{\dd t} \ket{\rho(t)} = \LL_+(t)\ket{\rho(t)},\qquad
\LL_+(t) = \un{\aaa}\cdot\mm{A}(t)\un{\aaa} - A_0(t)\hat{\one},
\label{eq:Liouvt}
\end{equation}
where explicit time-dependece of the structure matrix $\mm{A}(t)$ may physically arise due to driving by means of an external time-dependent force (time dependent matrix $\mm{H}$(t)) or time dependent coupling operators (time dependent vectors $\un{x}_\mu(t)$).
In this situation NESS cannot exist, but we shall show that one may still efficiently evaluate the propagator
\begin{equation}
\ket{\rho(t)} = \hat{\cal U}\ket{\rho(0)},\qquad \hat{\cal U}:= \hat{\cal T}\exp\left( \int_0^t \dd \tau \LL_+(\tau)\right),
\label{eq:tord}
\end{equation}
where $\hat{\cal T}$ indicates a time-ordered product.

The procedure is the following. Note that the space of all anti-symmetric complex structure matrices form a Lie algebra ${\rm so}(4n,\CC)$.
The following straightforward identity
\begin{equation}
[{\textstyle\frac{1}{2}}\un{\aaa}\cdot\mm{A}\un{\aaa},{\textstyle\frac{1}{2}}\un{\aaa}\cdot\mm{B}\un{\aaa}] = {\textstyle\frac{1}{2}}\un{\aaa}\cdot[\mm{A},\mm{B}]\un{\aaa},
\end{equation}
holding for any pair of complex $4n\times 4n$ matrices $\mm{A},\mm{B}$, indicates that Liouvilleans (\ref{eq:LiouvA},\ref{eq:Liouvt}) generate $4^n$ dimensional
representation of $so(4n,\CC)$. Thus, the time-ordered product (\ref{eq:tord}) can be evaluated within a Lie group $SO(4n,\CC)$ of $4n\times 4n$ matrices,
\begin{equation}
\mm{U} = \hat{\cal T}\exp\left(2\int_0^t\!\!\dd \tau \mm{A}(\tau)\right)
\end{equation}
and\footnote{Even if this has to be done numerically, using Trotter-Suzuki decomposition schemes, the computational compexity is only polynomial in $n$.}  then full Liouvillean propagator is written as
\begin{equation}
\hat{\cal U} = \exp(\un{\aaa}\cdot\mm{C}\un{\aaa} - C_0 \hat{\one}),\quad \mm{C} = \frac{1}{2}\ln\mm{U},\quad C_0 = \int_0^t\dd\tau A_0(\tau).
\end{equation} 
The logarithm of $\hat{\cal U}$ can be now considered as a `static' Liouvillean, so we can diagonalize it by the methods of subsection (\ref{subsect:static}),  leading to spectral decomposition of the form (\ref{eq:prop}).

\section{XY spin chains}

\label{sect:xy}

The theory of the previous two sections shall now be applied to investigate a homogeneous, finite XY chain of $n$ spins, described by Pauli matrices 
$\sigma^{\x,\y,\z}_j,j=1,\ldots n$ with the Hamiltonian
\begin{equation}
H =
\sum_{j=1}^{n-1} \left( { \frac{1+\gamma}{2}} \sigma^\x_j \sigma^\x_{j+1} + { \frac{1-\gamma}{2}} \sigma^\y_j \sigma^\y_{j+1}\right)
+ \sum_{j=1}^n h \sigma^\z_j,
\label{eq:hamsc}
\end{equation}
which is described by two real parameters, anisotropy $\gamma$ and transverse magnetic field $h$. Without loss of generality we may assume that $\gamma,h\in[0,\infty)$.
We decide to couple XY chain {\em thermally} only at its ends, so we consider the most general four coupling operators which allow for an explicit solution
\begin{eqnarray}
\label{sklop oper}
X_1&= \kappa_1 ( \sigma_1^\x\cos\theta_1+\sigma_1^\y\sin\theta_1),  
\quad 
X_3= \kappa_3 (\sigma_{N}^\x\cos\theta_3 + \sigma_N^\y\sin\theta_3), \nonumber \\
X_2&=\kappa_2(\sigma_1^\x\cos\theta_2+\sigma_1^\y\sin\theta_2),  \quad X_4=\kappa_4(\sigma_N^\x\cos\theta_4+\sigma_N^\y\sin\theta_4),
\label{eq:Xs}
\end{eqnarray}
and fully decorrelated baths $\Gamma_{\mu,\nu}^\beta = \delta_{\mu,\nu}\Gamma_\mu^\beta$. We take standard baths of harmonic oscillators at two ends with possibly different inverse
temperatures, and Ohmic spectral functions
\begin{equation}
\tilde{\Gamma}^{\beta_\mu}_{\mu,\nu}(\omega) = \lambda^2 \delta_{\mu,\nu} \frac{ \omega}{\exp(\omega \beta_\mu)-1}, \quad \beta_{1,2}\equiv \beta_\L,\quad \beta_{3,4}\equiv \beta_\R.
\label{eq:Ohmic}
\end{equation}
Note that frequency cutoff in the spectral function is irrelevant as we neglect the Lamb shift term in the master equation.

The enitre problem can be fermionized by means of Jordan-Wigner transformation (\ref{eq:jordan}), namely the Hamiltonian and the coupling operators transform to
\begin{eqnarray}
H&=&-\mathrm{i}\sum_{j=1}^{n-1}\Big(\frac{1-\gamma}{2}w_{2j}w_{2j+1}-\frac{1+\gamma}{2}w_{2j-1}w_{2j+2}\Big) - \mathrm{i}\sum^n_{j=1}h w_{2j-1}w_{2j}, \nonumber \\
X_1&=& \kappa_1( w_1\cos\theta_1+w_2\sin\theta_1), \;
X_3= W \kappa_3 \big(w_{2n}\cos\theta_3-w_{2n-1}\sin\theta_3\big), \label{eq:X}\\
X_2&=& \kappa_2 (w_1\cos\theta_2 + w_2\sin\theta_2), \; X_4= W \kappa_4 \big(w_{2n}\cos\theta_4-w_{2n-1}\sin\theta_4\big), \nonumber
\end{eqnarray}
where $W=(-\ii)^{n-1} w_1 w_2 \cdots w_{2n}$ is an operator which commutes with all the elements of ${\cal K}^+$ (or anti-commutes with all the elements of ${\cal K}^-$) and satisfies
$WW^\dagger = W^\dagger W = 1$, hence it has {\em no effect} on the dissipator (\ref{eq:redfD}) in $\LL_+$. We note however, that the commutation of $W$ thru $\rho$ in (\ref{eq:redfD}) for the dynamics in ${\cal K}^-$ produces a {\em minus} sign in all the bath terms, i.e. it changes the sign of $\PP_- \DD\PP_-$, with respect to a pure fermionic problem.

The $4n\times 4n$ structure matrix has now a specific block-tridiagonal + block-bordered form,
\begin{equation}
\mm{A} = \mm{A}' + \mm{B},
\label{eq:struct1}
\end{equation}
with
\begin{equation}
\!\!\!\!\!\!\!\!\!\!\!\!\!\!\!\!
\mm{A}' =
\pmatrix{
\mm{a} & \mm{b} & \mm{0} & \mm{0} & \ldots & \mm{0} \cr
\mm{c} & \mm{a} & \mm{b} & \mm{0} & \ldots & \mm{0} \cr
\mm{0} & \mm{c} & \mm{a} & \mm{b} & \ldots & \mm{0} \cr
\vdots & & \ddots & \ddots & \ddots &  \vdots \cr
\mm{0} & \mm{0} & \ldots & \mm{c} & \mm{a} & \mm{b} \cr
\mm{0} & \mm{0} & \ldots & \mm{0} & \mm{c} & \mm{a}},
\;
\mm{B} = \pmatrix{
\mm{l}_1 & \mm{l}_2 & \ldots & \mm{l}_{n-1} & \mm{l}_n \cr
\mm{l}'_2 & \mm{0} & \ldots  & \mm{0} & \mm{r}'_2 \cr
\vdots & \vdots & \ddots & \vdots & \vdots \cr
\mm{l}'_{n-1} & \mm{0} & \ldots & \mm{0} & \mm{r}'_{n-1} \cr
\mm{r}_1 & \mm{r}_2 & \ldots & \mm{r}_{n-1} & \mm{r}_n \cr
},
\label{eq:bb}
\end{equation}
where $\mm{a},\mm{b},\mm{c}$ are $4\times 4$ matrices
\begin{equation}
\mm{a}= -\ii h \one_2 \otimes \sigma^{\rm y}, \quad
\mm{b} = \frac{1}{2} \one_2 \otimes ( \ii \sigma^{\rm y} - \gamma \sigma^{\rm x}),
\quad
\mm{c} = -\mm{b}^T.
\label{eq:abc}
\end{equation}
The sequences of $4\times 4$ matrices $\mm{l}_j,\mm{l}'_j,\mm{r}_j,\mm{r}'_j$ which form the block-bordered part $\mm{B}$ can be straightforwardly computed [seeing (\ref{eq:explA})] from the form of the coupling vectors $\un{x}_{1,2} = (\kappa_{1,2}\cos\theta_{1,2},\kappa_{1,2}\sin\theta_{1,2},0,\ldots 0)^T$,
$\un{x}_{3,4} = (0,\ldots,0,-\kappa_{3,4}\sin\theta_{3,4},\kappa_{3,4}\cos\theta_{3,4})^T$, and their bath-transformations (\ref{eq:znu}) with (\ref{eq:Ohmic}).
Although we are unable to give closed form general expressions, we can make an asymptotic estimate - for large $n$ - on the decay of these matrices with their distance from the diagonal
\begin{equation}
||\mm{l}_j|| \sim ||\mm{l}'_j|| \sim ||\mm{r}_{n+1-j}|| \sim ||\mm{r}'_{n+1-j}|| \propto \exp(-K j).
\end{equation}
The coefficient $K > 0$ in general depends only on $\gamma,h$, and $\beta_{\rm L}$ (for $\mm{l}_j$) or  $\beta_{\rm R}$ (for $\mm{r}_j$). 
Note that for the special case of {\em local Lindblad coupling} (\ref{eq:lindD}) with the same local coupling operators (\ref{eq:X}) , the only non-vanishing blocks which remain are the diagonal ones $\mm{l}_1$ and $\mm{r}_n$, given explicitly in Ref.\cite{njp}.

Below we shall present some intriguing numerical results of the non-equilibrium thermal Redfield equation (\ref{eq:master},\ref{eq:redfD}) for the open XY chain given by (\ref{eq:hamsc},\ref{eq:Xs},\ref{eq:Ohmic}), in comparison with the local non-equilibrium Lindblad model (\ref{eq:lindD}) where a suitable set of coupling
operators of the form (\ref{eq:Xs}) and $4\times 4$ coupling matrix $\gamma_{\mu,\nu}$ can be chosen to parametrize the Lindblad operators 
$L_{1,2} = \sqrt{\Gamma^{\rm L}_{1,2}} \sigma^{\mp}_1$, $L_{3,4} =  \sqrt{\Gamma^{\rm R}_{1,2}} \sigma^{\mp}_n$,
parametrized exactly in the same way as in Refs.\cite{njp,pp08}.
For all the numerical results reported for the thermal Redfield model we consider the bath parameter values $\kappa_1=\kappa_3=1$, $\kappa_2=\kappa_4=0$, $\theta_1=\theta_3=\pi/6$,
and $\beta_{\rm L} = 0.3$, $\beta_{\rm R}=5.2$ unless $\beta$'s are varying, and $\lambda=0.1$ unless $\lambda$ is varying, whereas for the Lindblad model we always take the bath parameters
$\Gamma^{\L}_{1}=0.5$, $\Gamma^{\rm L}_2=0.3$, $\Gamma^{\rm  R}_1=0.5$, $\Gamma^{\rm R}_2=0.1$.

\subsection{Non-equilibrium phase transition}

\label{subsect:nept}

\begin{figure}[htb]
\begin{center}
\includegraphics[scale=0.44]{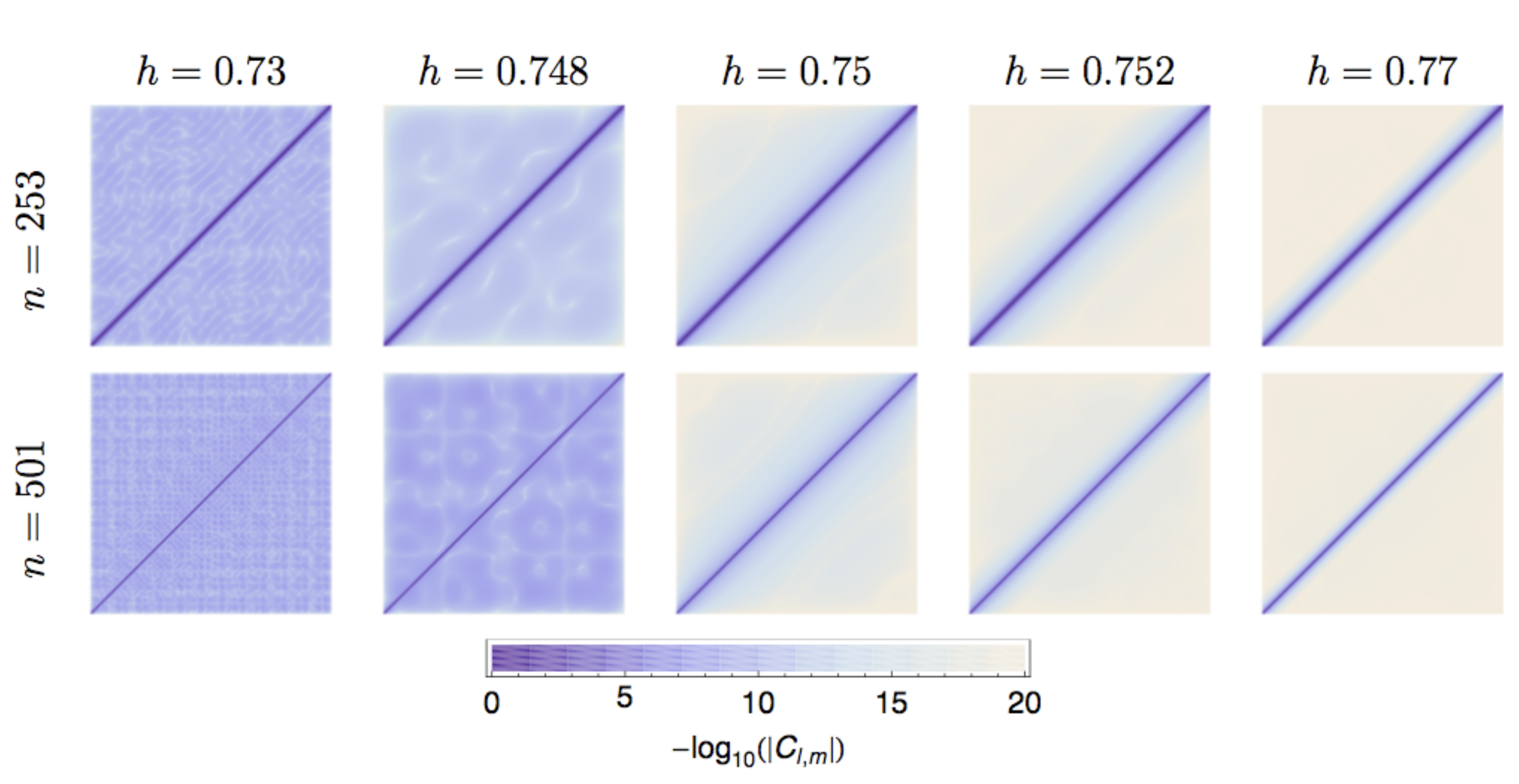}

\caption{Correlation matrices $C_{l,m}$, $l$ horizontal axis (left to right), $m$ vertical axis (bottom to top), of the non-equilibrium thermal Redfield model of an open XY chain
for $\gamma=0.5$ and different field strength $h$ indicated at the figures (note that $h_{\rm c}=0.75$) and two diffetrent system sizes $n$ (indicated). 
Bath parameters are specified in the text.}
\label{cormat}
\end{center}
\end{figure}

In Ref.\cite{pp08} an intriguing suggestion of a quantum phase transition far from equilibrium in the steady state of an open boundary driven XY spin chain has been put forward.
Numerical and heuristic theoretical evidence has been given for the spontaneous emergence of long range magnetic order in NESS as soon as the magnetic field
drops below the critical value $|h| < h_{\rm c}$,
\begin{equation}
h_{\rm c} = |1-\gamma^2|.
\end{equation} 
However, that study was done with local Lindblad reservoirs, so the questions remained whether the effect persists in the presence of local thermal reservoirs satisfying 
KMS conditions for non-vanishing temperatures. It is an easy task now to follow the recipes of subsection \ref{subsect:static} and numerically evaluate the spin-spin correlator
(note the use of Wick theorem as the spin-spin correlator is of fourth order in $w_j$):
\begin{eqnarray}
C_{l,m} &=& \tr(\sigma^\z_l \sigma^\z_m \rho_{\rm NESS}) - 
\tr(\sigma^z_l \rho_{\rm NESS})\tr(\sigma^z_m \rho_{\rm NESS}) \label{eq:corr}\\
&=& \ave{w_{2l-1} w_{2m-1}}_{\rm NESS} \ave{w_{2l}w_{2m}}_{\rm NESS} - \ave{w_{2l-1} w_{2m}}_{\rm NESS} \ave{w_{2l}w_{2m-1}}_{\rm NESS}.
\nonumber
\end{eqnarray} 
First, we use efficient prescription (\ref{eq:wwv}) to compute correlation matrices at non-equilibrium conditions $\beta_\L=0.3 \neq \beta_\R=5.2$ and plot them
for two different system sizes and five different values of $h$ around $h_{\rm c}$ in figure~\ref{cormat}.
Results look qualitatively identical to those for the Lindblad driving, even for other quantities that were investigated numerically in detail in \cite{pp08}.

\begin{figure}[htb]
\begin{center}
\includegraphics[scale=0.44]{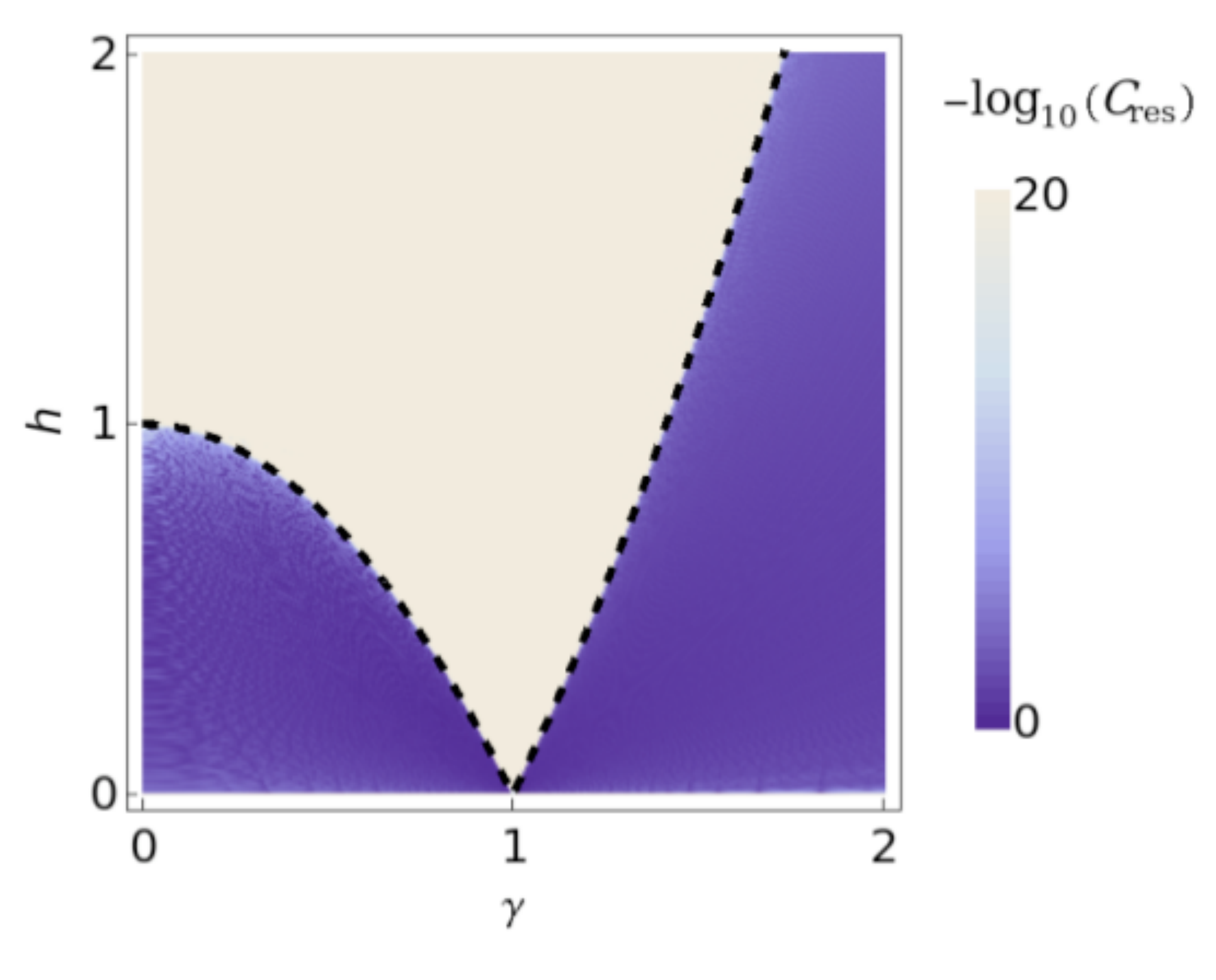}

\caption{Phase diagram for the non-equilibrium thermal Redfield model of an open XY chain. We plot the residual correlator $C_{\rm res}$ against the bulk parameters $\gamma$, $h$.
The system size is fixed to $n=100$ and bath parameters are specified in the text.
}
\label{phase}
\end{center}
\end{figure}

\begin{figure}[htb]
\begin{center}
	\includegraphics[scale=1]{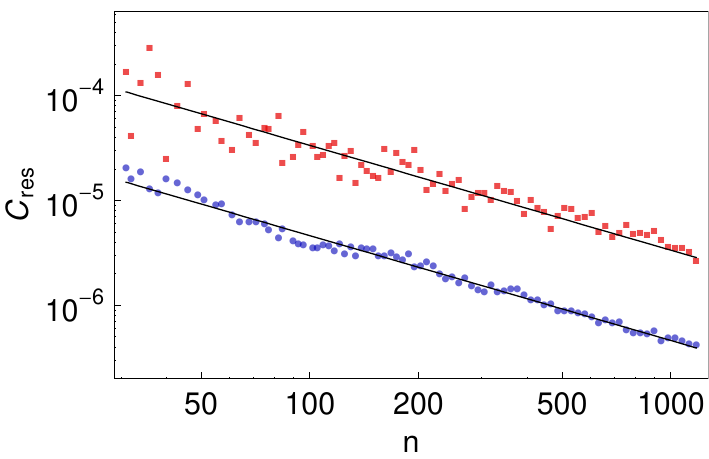}
	\includegraphics[scale=1]{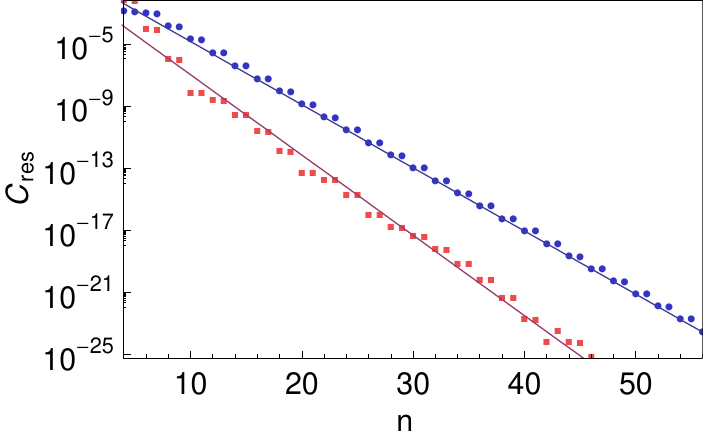}
\caption{Residual correlator $C_{\rm res}$ as a function of the system size $n$ for the LRMC phase ($\gamma=0.5, h=0.2$, left plot) and non-LRMC phase ($\gamma=0.5, h=0.9$, right plot), where we compare the non-equilibrium thermal Redfield model (red squares) and the non-equilibrum Lindblad model (blue circles) with bath parameters as specified in the text.
The thin lines indicated the suggested behavior $1/n$ (on the left) and $\exp(-\eta n)$ on the right 
(with the numerical best fit $\eta = 1.192$ for the Redfield model and $\eta = 0.937$ for the Lindblad model).}
\label{cresvsn}
\end{center}
\end{figure}

\begin{figure}[htb]
\begin{center}
	\includegraphics[scale=1]{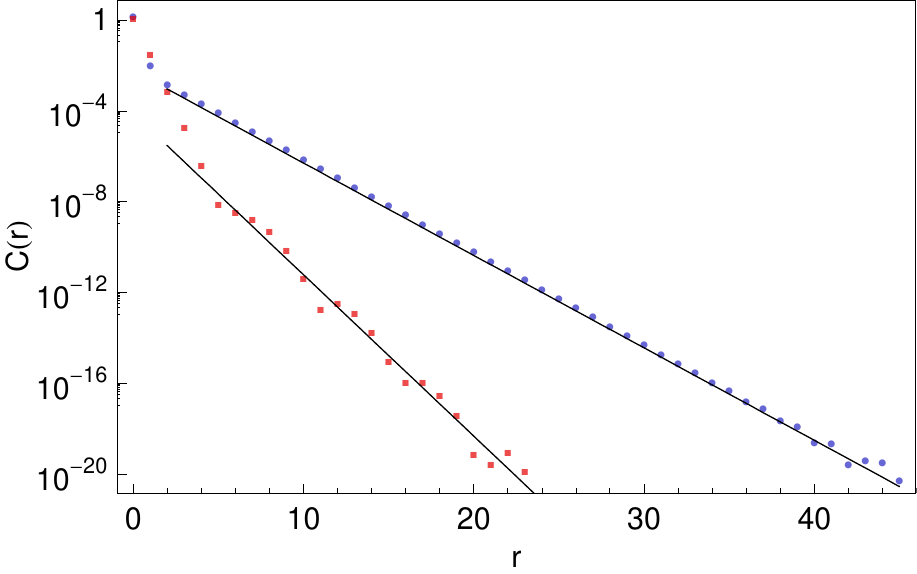}
\caption{Comparing the decay of the 2-point spin-spin correlator $C(r) = \sum_{j-i=r}C_{i,j}/\sum_{j-i=r} 1 \sim \exp(-\xi r)$ between the non-equlibrium thermal Redfield model (red squares)
and non-equilibrium Lindblad model (blue circles) for the same values of bulk parameters in the non-LRMC phase ($h=1.05, \gamma=0.2, n=200$) and bath parameters specified in the text.
The thin lines indicate suggested exponential decays $\propto \exp(-\xi r)$ with the exponents $\xi =1.635$ (fitting the Redfield model) and $\xi=0.937$ (fitting the Lindblad model).}
\label{profilwrld}
\end{center}
\end{figure}

For example, in figure~\ref{phase} we plot the phase diagram of the residual correlator 
$C_{\rm res} = \sum_{l,m}^{|l-m|>n/2}|C_{l,m}|/\sum_{l,m}^{|l-m|>n/2}1$, 
which also reveals possible criticality in the region of a large anisotropy $\gamma > 1$ previously not discussed.
We note that size dependence of the residual magnetic correlator $C_{\rm res}$ shows a very characteristic behaviour: namely
\begin{eqnarray}
C_{\rm res} &\propto& \exp(-\eta n)\;
{\rm with}\; \eta > 0\quad {\rm for}\quad |h| > h_{\rm c}\; {\rm or}\; h=0 \label{eq:scaling1}\\
C_{\rm res}&\propto& 1/n\qquad\qquad\qquad\qquad\, {\rm for}\quad 0< |h| < h_{\rm c} \label{eq:scaling2}
\end{eqnarray} 
Thus we shall refer to the regime with $0 < |h| < h_{\rm c}$ as {\em long range magnetic correlation} (LRMC) phase\footnote{Note, interestingly, that unlike for the local Lindblad driving\cite{pp08} the XX line
$\gamma=0$, $0<|h|<1$, also exhibits long range magnetic correlations for the thermal Redfield 
driving.}, the
regime with $|h| > h_{\rm c}$, or $h=0$, as non-LRMC phase, and the regime with $|h| = h_{\rm c}$ as {\em critical}. 
Scaling (\ref{eq:scaling1},\ref{eq:scaling2}) is illustrated in figure \ref{cresvsn}.
Exponential decay of the $C_{\rm res}(n)$ in non-LRMC phase (\ref{eq:scaling1}) is consistent with the exponential decay
of 2-point correlator with the distance between sites 
$C(r) = \sum_{j-i=r}C_{i,j}/\sum_{j-i=r} 1 \sim \exp(-\xi r)$, as can be qualitatively noted already in the figure \ref{cormat}.
However, we demonstrate in figure \ref{profilwrld} that the exponents $\xi$ could in principle be very different between the Redfield and local Lindblad
models. Futhermore, as for the Linbdlad model the exponents
$\xi$ and $\eta$ [of (\ref{eq:scaling1})] appear to be equal, for the 
Redfield model they don't seem to be simply related. 
Analytical estimation of these exponents
present a challenge for future theoretical work.

\begin{figure}[htb]
\begin{center}
	\includegraphics[scale=1]{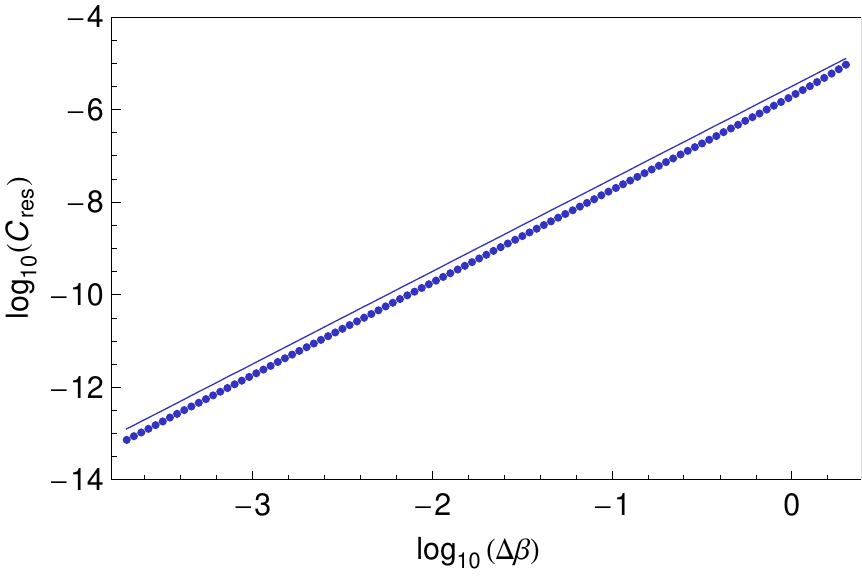}
\caption{Residual correlation $C_{\rm res}$ versus the (inverse) temperature drop $\Delta\beta$ between the left and the right bath, 
$\beta_L=2-\Delta \beta/2$, $\beta_R=2+\Delta \beta/2$, for the non-equilibrium thermal Redfield model of an open XY chain in LRMC phase
($\gamma=0.5, h = 0.3$), system size $n=100$, and the bath parameters specified in the text.
The thin line indicated suggested $|\Delta\beta|^2$ behavior.}
\label{fig:deltabeta}
\end{center}
\end{figure}

However, we note that with the thermal driving with Redfield dissipators, the long-range-magnetic order disappears when the temperatures of the baths become equal, $\beta_\L=\beta_\R$, and there we recover, consistently, all the properties of the thermal state \cite{mccoy} which are most easily 
numerically reproduced by the method of Ref.\cite{zpp08} , i.e. fast decay of correlations for any $h$ and absence of long-range order.
For example, it is interesting to note how the residual correlator $C_{\rm res}$ (for large $n$ in the LRMC phase) decreases as a function of the difference of inverse temperatures
$\Delta\beta=\beta_\R-\beta_\L$, namely numerics of figure \ref{fig:deltabeta} suggests clearly that  $C_{\rm res} \propto (\Delta\beta)^2.$

Heuristic explanation of this non-equilibrium phase transition is rather straightforward \cite{pp08}, however its exact proof and also the quantitative dependence of the decay exponent
$\eta(\gamma,h)$ are still lacking.
We note that the transition point $h=h_{\rm c}$ is characterized by a simple property of the XY spin chain quasiparticle dispersion relation
\begin{equation}
\omega(q) = \sqrt{(\cos q - h)^2 + \gamma^2 \sin^2 q}\ ,
\label{eq:xydisp}
\end{equation}
where $\epsilon_j=\omega(2\pi j/n)$ would be exactly the (positive) eigenvalues of matrix $\mm{H}$ if periodic boundary conditions would be imposed on the closed system.
Namely, in non-LRMC phase $|h| > h_{\rm c}$ there exist only a single pair of trivial stationary points $q^* = 0,\pi$, whereas in LRMC phase $|h| < h_{\rm c}$ there exist another pair of nontrivial stationary points $\pm q^* \neq 0,\pi$, $\dd\omega/\dd q|_{q=q^*} = 0$, which introduces a new non-trivial length scale $1/q^*$ which determines typical sizes of correlated regions in the matrix $C_{l,m}$ 
(see figure \ref{cormat}). Therefore this simple non-equilibrium quasi-particle picture predicts mean-field critical exponent $1/q^* \sim |h_{\rm c}-h|^{-1/2}$ as $h \uparrow h_{\rm c}$ 
(confirmed in Ref.\cite{pp08}).

\begin{figure}[htb]
\begin{center}
	\includegraphics[scale=1]{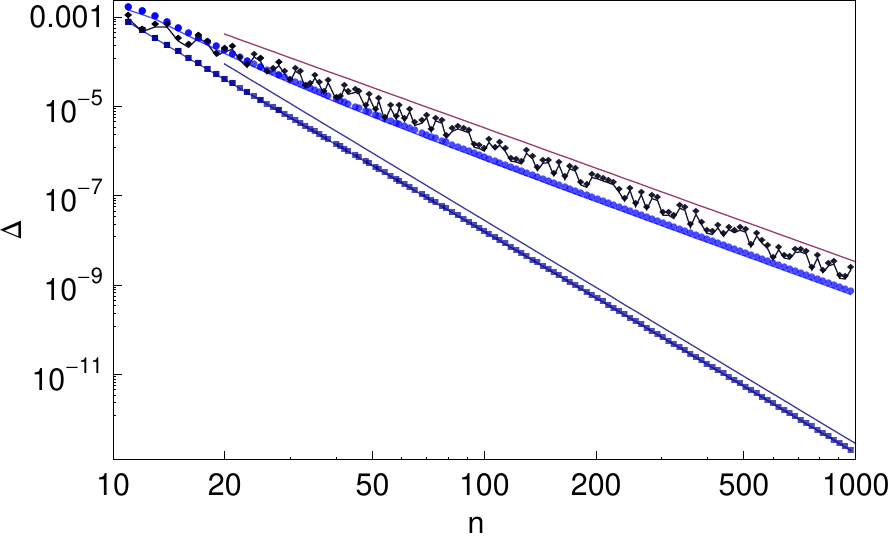}
\caption{Liouvilllean spectral gap $\Delta$ for the non-equilibrum thermal Redfield model of an open XY chain.
We plot three different cases with: $\gamma=0.5$, $h = 0.8 > h_{\rm c}$ (non-LRMC phase, light blue circles),
$\gamma=0.5$, $h=0.75=h_{\rm c}$ (critical regime, dark blue squares), 
$\gamma=0.5$, $h=0.3 < h_{\rm c}$ (LRMC phase, black diamonds), whereas the bath parameters are specified in the text.
Suggested power law decays $n^{-3}$ and $n^{-5}$ are indicated with thin lines.
}
\label{fig:gap}
\end{center}
\end{figure}

The non-equilibrium quantum phase transition can also be characterized by the scaling of the Liouvillean spectral gap $\Delta(n)$, namely in the
critical regime one expects a qualitative increase in the relaxation time $1/\Delta$ to NESS.
Numerical results (see figrure \ref{fig:gap}) suggest the spectral gap of the Liouvillean remains like in the local Lindblad case \cite{njp}
\begin{equation}
\Delta \propto n^{-3} \; {\rm for}\; h\neq h_{\rm c},\qquad \Delta \propto n^{-5} \; {\rm for}\; h = h_{\rm c},
\end{equation}
although we are at the moment unable to prove this conjecture. Also note slight fluctuations of $\Delta(n)$ in the LRMC phase as
opposed to a smooth power law in the non-LRMC phase.

\begin{figure}[htb]
\begin{center}
	\includegraphics[scale=1]{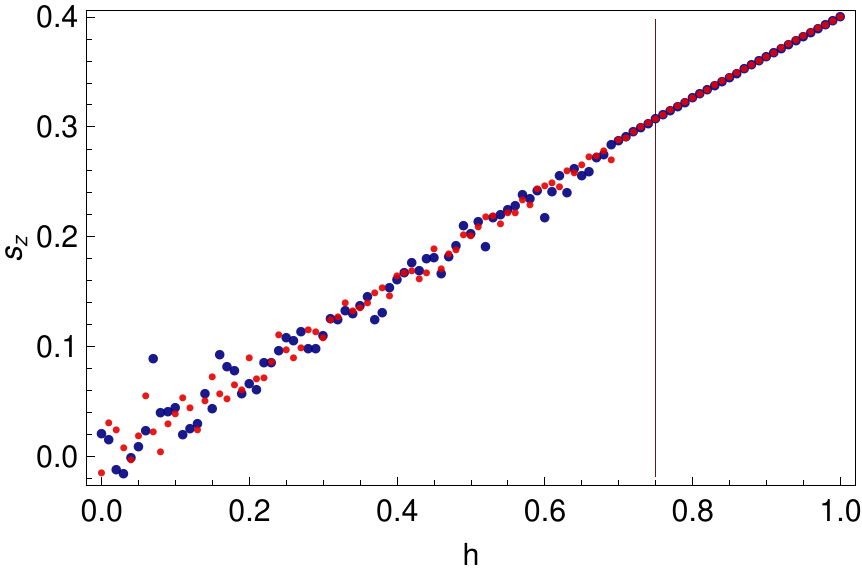}
\caption{Hypersensitivity of NESS to magnetic field strength $h$. We plot local magnetization $s_{\rm z}(h) = \ave{\sigma^{\rm z}_{n/2}}_{\rm NESS}$ for the non-equilibrium thermal
Redfield model of open XY chain with $\gamma=0.5$ and bath parameters as written in the text. Big blue (small red) circles represent data for $n=50$ ($n=100$), whereas vertical line denotes the critical value $h=h_{\rm c}$.}
\label{sensitivity}
\end{center}
\end{figure}

Long range correlations for $|h| < h_{\rm c}$ naturally imply sensitivity of NESS to tiny variations in system's parameters. For example, one may expect also that local observables in NESS will be
then sensitive functions of the bath-driving or even bulk parameters, such as the magnetic field $h$. In figure~\ref{sensitivity} we plot local magnetization in the center of the chain
$s_{\rm z} = \ave{\sigma^{\rm z}_{n/2}}_{\rm NESS}$ versus the field strength $h$. Indeed, we notice that for $|h| > h_{\rm c}$, $s_{\rm z}(h)$ is a {\em smooth} function wheres for
$|h| < h_{\rm c}$, $s_{\rm z}(h)$ becomes rapidly oscillating or better to say, fluctuating, function. Even though the amplitude of these oscillations decreases with $n$, the scale of $h$ on which 
$s_{\rm z}(h)$ fluctuates decreases with $n$ even much faster, so we predict that in the thermodynamic limit $n\to\infty$, in LRMC phase the local susceptibility
$\dd s_{\rm z} /\dd h$ would be {\em ill defined}.
In summary, LRMC phase can be characterized by {\em hypersensitivity of NESS to external parameters}.

\subsection{Heat transport and entropies}

An important non-equilibrium physical effect which one can investigate more deeply in an open XY chain is the {\em heat transport},
which has been recently intensively studied in quantum spin chains, 
see e.g.\cite{saito,fabian,mejia05,hartmann,aligia} or \cite{dhar} for a recent review on the topic.
 
\begin{figure}[htb]
\begin{center}
	\includegraphics[scale=0.44]{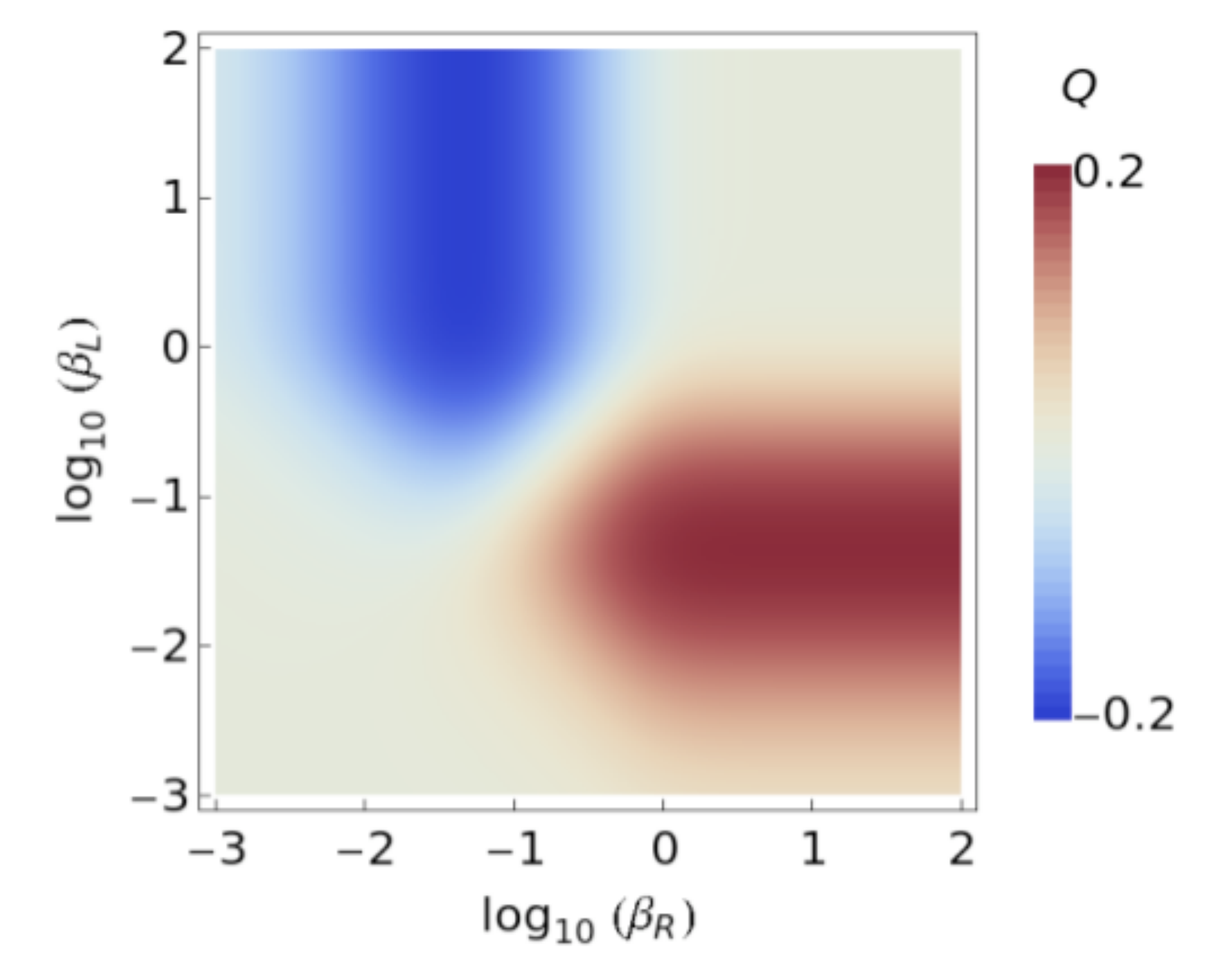}
\caption{NESS expectation value of the heat current $\ave{Q_m}_{\rm NESS}$ versus two inverse temperatures $\beta_{\rm L}$ and $\beta_{\rm R}$ for the 
non-equilibrium thermal Redfield model of an open XY chain with $\gamma=0.5, h=0.9$, sistem size $n=53$, and bath parameters given in the text. 
Note that the `shoulders' of maxima, around $\beta_{\rm L} \approx 0.05, \beta_{\rm R} > 1$, and with ${\rm L}$ and ${\rm R}$ exchanged, 
could be interpreted as {\em negative differential heat conductance}.}
\label{fig:tok}
\end{center}
\end{figure}

Writing the Hamiltonian (\ref{eq:hamsc}) in the bulk 
as a sum $H = \sum_m H_m$ with a two body energy density operator
\begin{eqnarray}
H_m &=&  -\ii \frac{1+\gamma}{2} w_{2m} w_{2m+1} + \ii \frac{1-\gamma}{2} w_{2m-1}w_{2m+2} \nonumber \\
&& -\ii\frac{h_m}{2}w_{2m-1}w_{2m}-\ii\frac{h_{m+1}}{2}w_{2m+1}w_{2m+2},
    \label{eq:hdens}
\end{eqnarray}
one can derive the local {\em energy current} 
\begin{eqnarray}
Q_m&=&\ii[H_m,H_{m+1}] \\
&=& \ii(1-\gamma^2) (w_{2m-1}w_{2m+3} +  w_{2m}w_{2m+4}) \nonumber\\
&-& 2\ii h(1-\gamma)(w_{2m-1}w_{2m+1} + w_{2m+2}w_{2m+4}) \nonumber\\
&-& 2\ii h(1+\gamma)(w_{2m} w_{2m+2}+w_{2m+1}w_{2m+3}), \nonumber
  \label{eq:hcurr}
  \end{eqnarray}
which, by construction, satisfies the continuity equation
\begin{equation}
(\dd/\dd t)\ave{H_m} = \ave{\ii[H,H_m]} + \tr H_m \hat{\cal D}\rho(t) = -\ave{Q_{m}} + \ave{Q_{m-1}}.
\label{eq:cont}
\end{equation}
The two terms between the two equality signs above correspond to the unitary and dissipative term in the master equation (\ref{eq:master}). The unitary term has been already transformed to
a simple expectation value using cyclicity of the trace $\tr x[y,z] \equiv \tr y[z,x]$, while the dissipative term can be further shown to vanish in the bulk $ 2 \le m \le n-2$ by 
excercising the cyclicity of the trace again and transforming the integrand of (\ref{eq:redfD}) to terms of the form $\tr \tilde{X}_\mu(-\tau)\rho [X_\nu,H_m] \equiv 0$. The RHS expression of eq. 
(\ref{eq:cont}) then
follows from the nearest-neighbour locality of the Hamiltonian.
Therefore, in NESS the expectation value of the current $\ave{Q_m}_{\rm NESS}$ should be independent of the position $m$. By looking at the dependence of the steady-state current on the system size we clearly find {\em ballistic transport}, namely $\ave{Q_m}_{\rm NESS} = {\cal O}(n^0)$, irrespectively of the temperature differences between the baths and bulk parameters of the model (i.e. whether being in the LRMC phase, non-LRMC phase, or critical).
However, we find very interesting dependence of the heat current on the temperature driving, i.e. on the two temperatures of the thermal baths.
In figure \ref{fig:tok} we plot $\ave{Q_m}_{\rm NESS}$ versus $\beta_{\rm L}$ and $\beta_{\rm R}$ and find a maximum of the current for intermediate driving, namely when one of the
temperatures is less than one $1/\beta_{\rm L} < 1$ and the other temperature is about $1/\beta_{\rm R} \approx 20$. This is a clear signature of {\em negative differential heat conductance} which could perhaps be related to similar far-from-equilibroum effects recently observed in spin and charge transport \cite{itaslo}.

\begin{figure}[htb]
\begin{center}
	\includegraphics[scale=0.44]{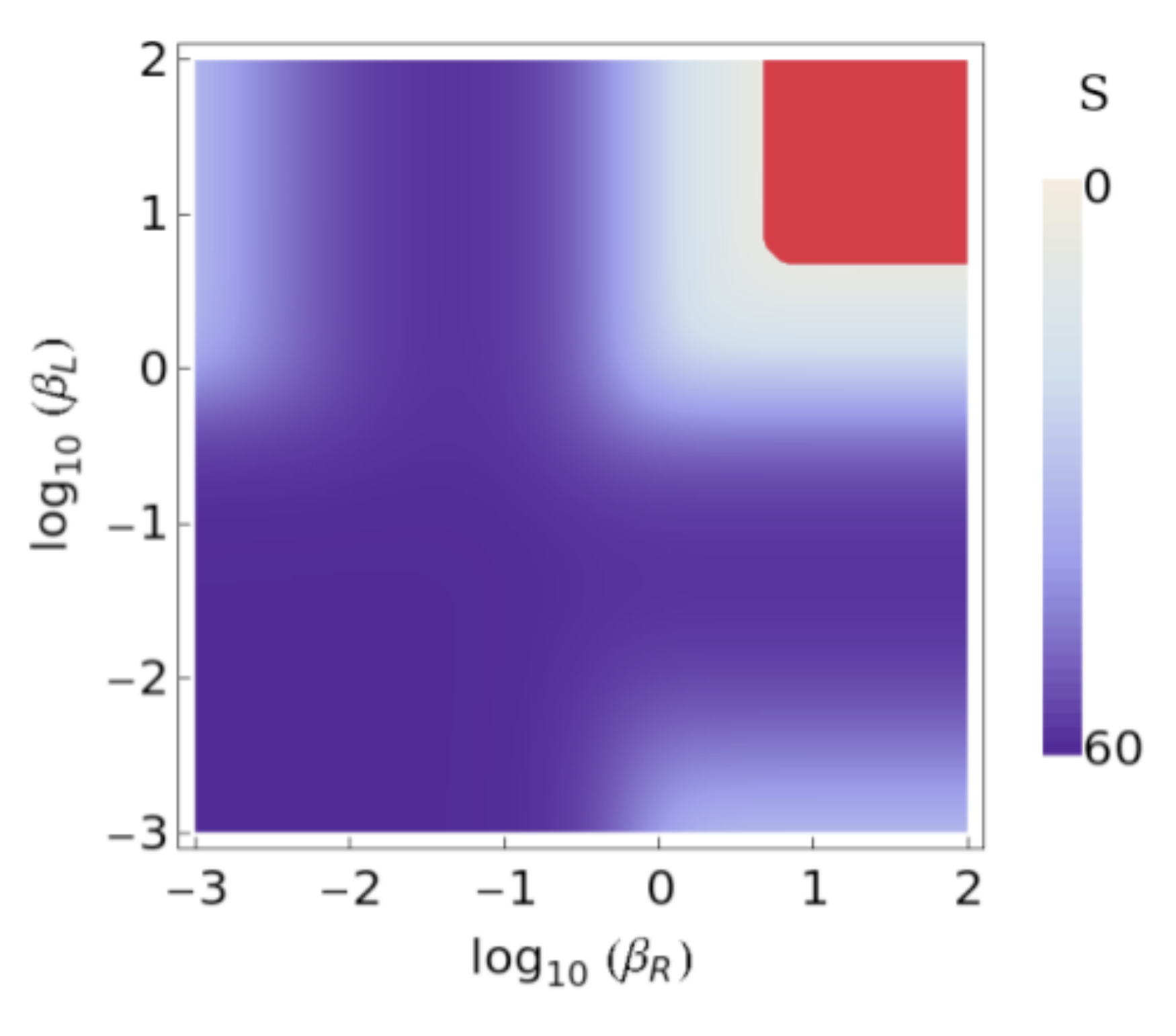}
\caption{Gibbs entropy of NESS versus two inverse temperatures $\beta_{\rm L}$ and $\beta_{\rm R}$ for the same parameters as in the figure \ref{fig:tok}.
Note that in the red region (of both large inverse temperatures),  
NESS is no longer a density matrix (at least one of the eigenvalues becomes slightly negative) hence the Gibbs entropy is strictly no longer defined there.}
\label{fig:entropy}
\end{center}
\end{figure}

This behavior can be nicely characterized by computing the Gibbs entropy of NESS. Since NESS is completely characterized by quadratic correlations $\ave{w_j w_k}_{\rm NESS}$ and the Wick theorem, one can adopt the recipe which has been proposed in Ref.\cite{latorre} for computing block entropies (or entanglement entropies) applied to the entire lattice.
In fact, taking an arbitrary block region $A \subseteq \{1,\ldots, n\}$,  one can compute Von Neumann entropy $S_A(\rho) = -{\rm tr}_{A} \rho_A \log_2 \rho_A$ (in base $2$), where $\rho_A = {\rm tr}_{\bar{A}} \rho$ is a reduced density matrix and $\bar{A}$ denotes the complement of $A$, as 
\begin{equation}
S_A = \sum_{j=1}^{ \#(A)} H_2((1+\nu_j)/2)
\; \;{\rm with}\;\;
H_2(x) := -x \log_2 x - (1-x)\log_2 x
\end{equation} 
and $\pm \ii \nu_j$ are the eigenvalues of the $2 \#(A) \times 2 \#(A)$ part  of the correlation matrix $B_{j,k}$ defined by 
$\ave{w_j w_k}_{\rm NESS}  =: \delta_{j,k} + \ii B_{j,k}$, restricted to Majorana operators $w_j, w_k$ corresponding to spins from the block $A$.
The same general procedure has been applied to thermal (Gibbs) states in Ref.\cite{zpp08}.
When taking the maximal block $A=\{1,\ldots,n\}$ we obtain exactly the standard Gibbs entropy of NESS. In figure \ref{fig:entropy} we plot the Gibbs entropy $S_{\{1,\ldots,n\}}$ as a function of two bath temperatures and show that, quite remarkably, the regions of large (maximal) heat current correspond to regions of large (locally maximal) Gibbs entropy. This is not unexpected as the product of the heat current and the inverse temperature difference $\Delta\beta$ may be understood as the {\em entropy production rate}. 

Calculation of Gibbs entropy of NESS provides also a nice way of controlling the positivity of NESS as
a density matrix, since this is by no means guaranteed by the Redfield master equation. Indeed we find that for very small temperatures (large $\beta$'s), or for very strong bath coupling $\lambda$, the positivity of NESS might be slightly violated (red region in figure \ref{fig:entropy}), namely some of the correlation matrix eigenvalues $\nu_j$ become slightly larger than $1$ (but in our numerical experience never by more than $10^{-7}$ or so).

We can use the concept of block entropy of NESS to further characterize the non-equilibrium phase transition.
For example, we may compute the total (quantum plus classical) correlations between two halves of the spin chain in NESS as given by {\em quantum mutual information} QMI
$I(n) = S_{\{1,\ldots,n/2\}} + S_{\{n/2+1,\ldots,n\}} - S_{\{1,\ldots,n\}}$.

\begin{figure}[htb]
\begin{center}
	\includegraphics[scale=1]{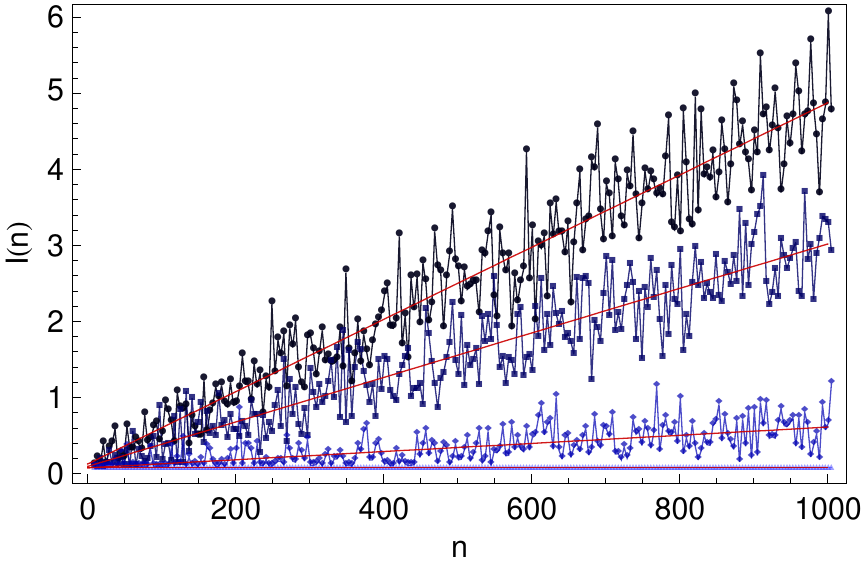}
\caption{
Another manifestation of the non-equilibrium phase transition: Quantum mutual information (QMI) of NESS for non-equlibrium thermal Redfield model of open XY spin chain.
The bulk parameters are $\gamma=0.5$ and $h=0.9 > h_{\rm c} = 0.75$ (lightest blue, saturated curve),
$h = 0.7$, $h=0.5$ and $h=0.3$ (from lighter to darker blue curves). Thin red lines indicated the linear growth of QMI for $h < h_{\rm c}$.}
\label{fig:qmi}
\end{center}
\end{figure}

Interestingly, we find (see figure \ref{fig:qmi}) that QMI saturates $I(n) = {\cal O}(n^0)$ in the non-LRMC phase (for $|h| > h_{\rm c}$), whereas in LRMC phase (for $0 < |h| < h_{\rm c}$) QMI becomes extensive $I(n) = {\cal O}(n)$ indicating a drastic enhancement of correlations in NESS. This is again very similar to the behaviour of {\em operator space entanglement entropy}  (OSEE)
(analized for the Lindblad model in \cite{pp08}), so one may extend the relationship between QMI and OSEE which has been conjectured for thermal states in Ref.\cite{zpp08} to NESS.

\begin{figure}[htb]
\begin{center}
	\includegraphics[scale=1]{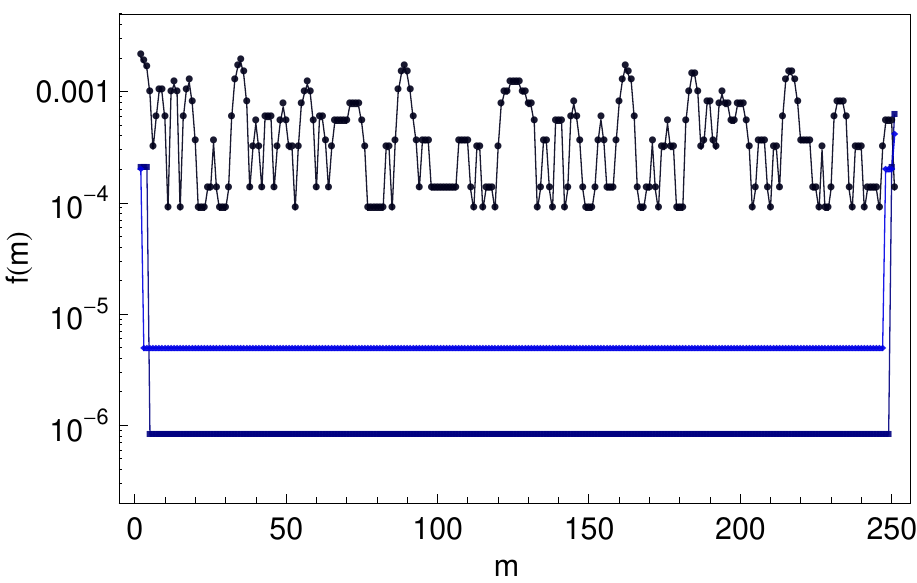}
\caption{Another manifestation of non-equilibrium phase transition: positional fluctuations in energy density in NESS of non-equilirbium thermal Redfield model
of open XY chain. We plot the relative fluctuation $f(m)=|\ave{H_m}_{\rm NESS} - \bar{H}|/|\bar{H}|$ where $\bar{H}$ is the bulk average of 
energy density $\ave{H_m}_{\rm NESS}$. Three curves correspond to $\gamma=0.5$ and $h=0.7 < h_{\rm c}$ (black curve), $h=0.75=h_{\rm c}$ (dark blue curve) and
$h=0.8 > h_{\rm c}$ (light blue curve), while the system size is $n=253$.
}
\label{density}
\end{center}
\end{figure}

In the context of energy transport it is interesting to look at the energy density profiles in NESS. In figure \ref{density} we plot the relative
spatial fluctuation of the energy density $f(m)=|\ave{H_m}_{\rm NESS} - \bar{H}|/|\bar{H}|$ where  $\bar{H} = (n-3)^{-1}\sum_{m=2}^{n-2} \ave{H_m}_{\rm NESS}$ is the averaged
energy density. Quite strikingly, we observe a big variation of $f(m)$ from site to site for LRMC phase and very smooth (non-fluctuating) behaviour for the non-LRMC phase which is 
characterized with a bulk-constant $f(m)$ which is exponentially small in $n$. This behaviour can again be considered as a manifestation of hypersensitivity of NESS and LRMC.

\begin{figure}[htb]
\begin{center}
	\includegraphics[scale=1]{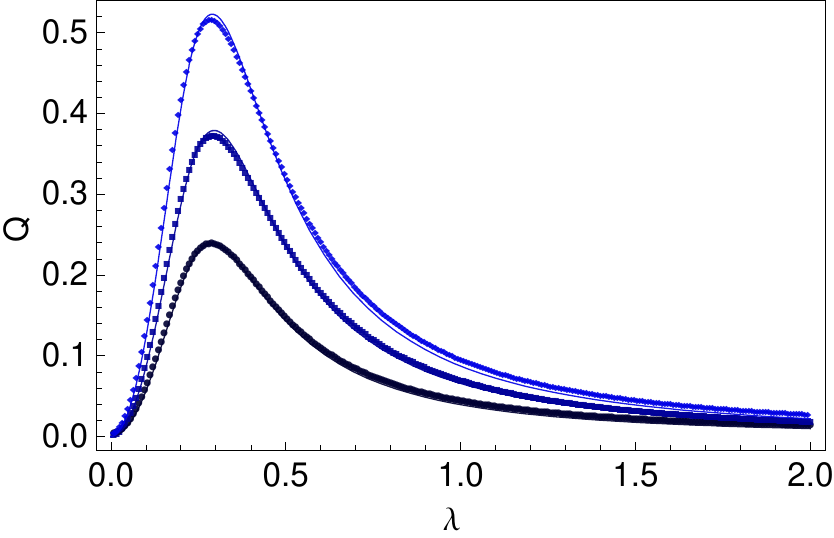}
\caption{NESS expectation value of the heat current $\ave{Q_m}_{\rm NESS}$ versus the coupling strength $\lambda$ for the 
non-equilibrium thermal Redfield model of an open XY chain with $h=0.5 < h_{\rm c}$ (black curve), $h=0.75=h_{\rm c}$ (dark blue curve) and
$h=1.0 > h_{\rm c}$ (light blue curve), system size $n=200$ and other bath parameters as given in the text.
Note that the full line gives numerically excellent fit to the Karevski and Platini formula \cite{karevski}
$\ave{Q_m} = a \lambda^2/ (b + \lambda^4)$ for best fitted parameters $a=0.040,b=0.0070$, $a=0.066,b=0.0076$, and $a=0.088,b=0.0071$ for the three cases of
$h=0.5,0.75,1.0$, respectively.  
}
\label{karevski}
\end{center}
\end{figure}

At last we check the dependence of the heat current $\ave{Q_m}_{\rm NESS}$ on the system-bath coupling strength $\lambda$.
It was recently reported by Karevski and Platini \cite{karevski} that the {\em spin current} $J_m$ in the local Lindblad model of an open isotropic XX chain $\gamma=0$ has a non-monotonic dependence on $\lambda$ which can be universally described by a formula $\ave{J_m}_{\rm NESS} = a'\lambda^2/(b' + \lambda^4)$ where $a',b'$ are some constants. 
For the anisotropic XY model and general non-equilibrium thermal Redfield driving we are unable to derive an exact analytic result, however our numerical simulations suggest a very 
similar behaviour for the heat current
\begin{equation}
\ave{Q_m}_{\rm NESS} \approx a\lambda^2/(b + \lambda^4),
\label{eq:karformula}
\end{equation}
where $a,b$ are again some constants which may depend on all system's parameters except $\lambda$. This is particulary interesting as in the anisotropic XY model the spin current is not even well
defined as there is no corresponding conservation law.
This behaviour is demonstrated in figure \ref{karevski} where on may also notice small but detectable deviations between numerics and the best fit to (\ref{eq:karformula}).
We note that the error of the fit does not decrease but is roughly constant when we increase the system size $n$.

\section{Discussion}

The purpose of the present paper was three-fold. Firstly, we have outlined a general method for exact treatment of quadratic many-body Markovian master equations. 
Our formalism, which rests upon treating density operators as elements of a suitable operator Fock space (or Liouville-Fock space) is quite flexible and allows for explicit solution of static 
and time-dependent quantum many-body Liouvillean problems, for example computation of arbitrary physical obsevables in the non-equilibrium steady state, decay rates of approach to the 
steady state, or even time-evolution of the density matrix of externally forced systems described by explicitly time-dependent Liouvilleans, all with polynomial computation complexity in number of particles (fermionic degrees of freedom).

Secondly, we have analyzed in detail the Redfield model of thermal baths within our framework.
In spite of the fact that the Redfield model does not define a proper dynamical semigroup, namely it is not guaranteed to preserve positivity of the density operator,
we have confirmed that steady states typically correspond to proper (positive) density operators. Tiny deviations from positivity have only been observed in some test cases for very small
temperatures or very large couplings to the baths (which anyway violate {\em weak coupling assumption}).
Furthermore, we have shown that coupling the central system with several thermal baths of the Redfield type at different temperatures produces physically interesting non-equilibrium steady states, for example such states which carry non-vanishing heat current. We wish to stress this physically obvious but mathematically delicate point with a particular care, as we have found a qualitatively different result for Lindblad-Davies dissipators which generate proper dynamical semigroups and satisfy detailed balance condition with respect to Gibbs states \cite{davies,alicki}.
Namely when we constructed a Lindblad-Davies dissipator with respect to two baths with two different temperatures coupled to two ends of the system (spin chain), we have found that the
resulting steady state (fixed point of the Liouvillean dynamics) is simply some convex combination of two Gibbs states corresponding to the bath temperatues, and as such has always zero
heat current and cannot represent physical steady state. This implies that the {\em secular approximation} (sometimes called the rotating wave approximation) which is the one-step from the Redfield to the Lindblad-Davies bath model 
prohibits the emergence of the physical out-of-equilibrium steady states with currents, therefore the seemingly harmless rapidly oscillating terms in the Redfield dissipator may be absolutely essential for non-equilibrium physics.
Thus we conjecture that the thermal Redfield model is somehow a minimal mathematical model which can describe non-equilibrium
thermal driving of a (non-self-thermalizing, e.g. integrable) open quantum system. 

Thirdly, we have applied our theory to analyze non-equilibrium quanutm phase transition and heat transport in an open XY spin 1/2 chain. We have carefully compared numerical results for the
non-equilibrium thermal Redfield model and the local Lindblad model, which has been discussed before \cite{njp,pp08}.
We have found that the phase diagram of the non-equilibrium XY model is insensitive to the theory with which we describe the baths, and the differences were only quantitative.
In particular we wish to stress that thermally driven heat current in the XY chains exhibits non-monotonic dependence on the temperature difference which may be interpreted as {\em negative differential heat conductance}. We believe that our numerical results on non-equilibrium open XY chain provide a strong motivation for further analytical work. In particular, we believe that the
block-tridiagonal plus block-bordered structure of the Liouvillean structure matrix (\ref{eq:struct1},\ref{eq:bb}) could be explored in combination with the non-equilibrium Green function formula
for the observables (\ref{eq:wwg},\ref{eq:Green}) to yield explicit asymptotic results for large $n$.  

Note added: Formally quite similar approach to non-equilibrium quasi-particles
has recently been developed independently by Kosov \cite{kosov}.

We acknowledge financial support by the Programme P1-0044, and the Grant  J1-2208, of the Slovenian Research Agency (ARRS).

\end{document}